\documentclass[aps,prb,twocolumn,superscriptaddress]{revtex4-1}

\usepackage[T1]{fontenc}
\usepackage{amsmath}
\usepackage{amssymb}
\usepackage{wasysym}
\usepackage{graphicx}
\usepackage{xcolor}
\usepackage{graphicx}
\usepackage{braket}
\usepackage{multirow}
\usepackage{soul}
\usepackage[normalem]{ulem}
\usepackage{placeins}

\usepackage[linktocpage=true,
  colorlinks=true, 
  pdfborder={0 0 0},
  linkcolor=blue,
  citecolor=blue,
  filecolor=yellow,
  urlcolor=blue,
  bookmarks,
  pdfauthor={},
]{hyperref}

\bibpunct{[}{]}{,}{n}{}{}

\newcommand{\Graz}{Institute of Theoretical and Computational Physics, Graz University of Technology, NAWI Graz, 8010 Graz, Austria}
\newcommand{\Rome}{Dipartimento di Fisica, Sapienza Universit\`a di Roma, 00185 Roma, Italy}
\newcommand{\Oxford}{Department of Materials, University of Oxford, Parks Road, Oxford OX1 3PH, United Kingdom}
\newcommand{\tc}{$T_\text{c}$}

\newcommand{\SC}{\textit{$Pm$\={3}$m$}}

\definecolor{mygreen}{rgb}{0.0, 0.6, 0.0}
\begin{document}

\title{Absence of superconductivity in iron polyhydrides at high pressures}

\author{Christoph Heil} \email{christoph.heil@tugraz.at}
\affiliation{\Oxford} \affiliation{\Graz}
\author{Giovanni B. Bachelet}            
\affiliation{\Rome}
\author{Lilia Boeri}
\affiliation{\Rome}

\date{\today}

\begin{abstract}
  Recently, C. M. P\'epin \textit{et al.} [\href{http://science.sciencemag.org/content/357/6349/382}{Science \textbf{357}, 382 (2017)}] reported the formation of several new iron polyhydrides FeH$_x$  at pressures in the megabar range, and spotted FeH$_5$, which forms above 130 GPa, as a potential high-\tc \ superconductor,  because of an alleged layer of dense metallic hydrogen. Shortly after, two studies by A.~Majumdar \textit{et al.} [\href{http://dx.doi.org/10.1103/PhysRevB.96.201107}{Phys. Rev. B \textbf{96}, 201107 (2017)}] and A.~G.~Kvashnin \textit{et al.} [\href{http://dx.doi.org/10.1021/acs.jpcc.8b01270}{J. Phys. Chem. C \textbf{122}, 4731 (2018)}] based on {\em ab initio} Migdal-Eliashberg theory seemed to independently confirm such a conjecture. We conversely find, on the same theoretical-numerical basis, that neither FeH$_5$ nor its precursor, FeH$_3$, shows any conventional superconductivity and explain why this is the case. We also show that superconductivity may be attained by transition-metal polyhydrides in the FeH$_3$ structure type by adding more electrons to partially fill one of the Fe--H hybrid bands (as, e.g., in NiH$_3$). Critical temperatures, however, will remain low because the $d$--metal bonding, and not the metallic hydrogen, dominates the behavior of electrons and phonons involved in the superconducting pairing in these compounds.
\end{abstract}


\maketitle

\section{Introduction}
The discovery of a record superconducting \tc \ of 203~K in H$_3$S has confirmed Ashcroft's 15-year-old suggestion that hydrogen-dominant metallic alloys are good candidates for the high-\tc \ conventional superconductivity at lower pressure  than the one needed to turn molecular hydrogen into a metallic superconductor (as predicted by Ashcroft 50 years ago)~\cite{DrozdovEremets_Nature2015,PhysRevLett.21.1748,PhysRevLett.92.187002}. This has ignited an intense theoretical and experimental search for new superconductors at high pressures, and presently, three groups of materials seem to stand out: $(i)$ hydrogen itself, in its high-pressure molecular (insulating) and atomic (metallic) phases~\cite{PhysRevLett.100.257001,PhysRevB.84.144515,PhysRevB.93.174308}; $(ii)$ covalent hydrides, which form molecular solids at normal pressure and crystalline metals at high pressure~\cite{Flores-Livas2016,Duan_SciRep2014,PhysRevLett.114.157004,
Drozdov_PH3_arxiv2015,Flores_PH3_PRBR2016,shamp_decomposition_2015,Fu_Ma_pnictogenH_2016}; and $(iii)$ heavy-metal hydrides with open hydrogen cages (involving  Ca, Y, La, U, etc.), which form at $P \gtrsim 200$ GPa~\cite{Wang_PNAS2012_CaH6,Kruglov_UH,Liu_PNAS_la_hydrides}.
All of them are unusual metals with strongly directional 
bonds~\cite{Heil-Boeri_PRB2015}, yielding a large electron-phonon ($e$-ph) coupling, but each requires extreme stabilization pressures. Hence, the challenge in this field of research is to devise chemical strategies to obtain lower and lower formation pressures for such high-\tc \ conventional superconductors~\cite{Flores_H2O} by identifying (or ruling out) classes of plausible candidates and electronic bands relevant for their $e$-ph interaction.
\newline \indent
Recently, P\'epin \textit{et al.}~\cite{Pepin_FEH_2017} reported the synthesis of a new iron hydride at 130~GPa: the hydrogen-rich layered crystal FeH$_5$, a metal which seems like a promising candidate for high-\tc \ superconductivity; shortly after, two theoretical papers argued that FeH$_5$ should indeed exhibit \tc's as high as 56~K~\cite{Majumdar_FEH,Oganov_FEH}.
In this work, using {\em ab initio} Migdal-Eliashberg theory as implemented
in the \textsc{epw} code~\cite{ponce_epw:_2016}, we show that the  \tc \ of FeH$_5$ is actually $\leq 1$~K and that for this compound the picture of a dense two-dimensional metallic-hydrogen layer is not sensible; on the contrary, FeH$_5$ bears a very strong resemblance to its precursor, FeH$_3$, which is also not superconducting.
In both compounds the conductivity is dominated by \mbox{$d$--metal} bonding, which, under appropriate circumstances, may yield conventional superconductivity
but not high \tc.
\begin{figure}[t]
\includegraphics[width=0.95\columnwidth]{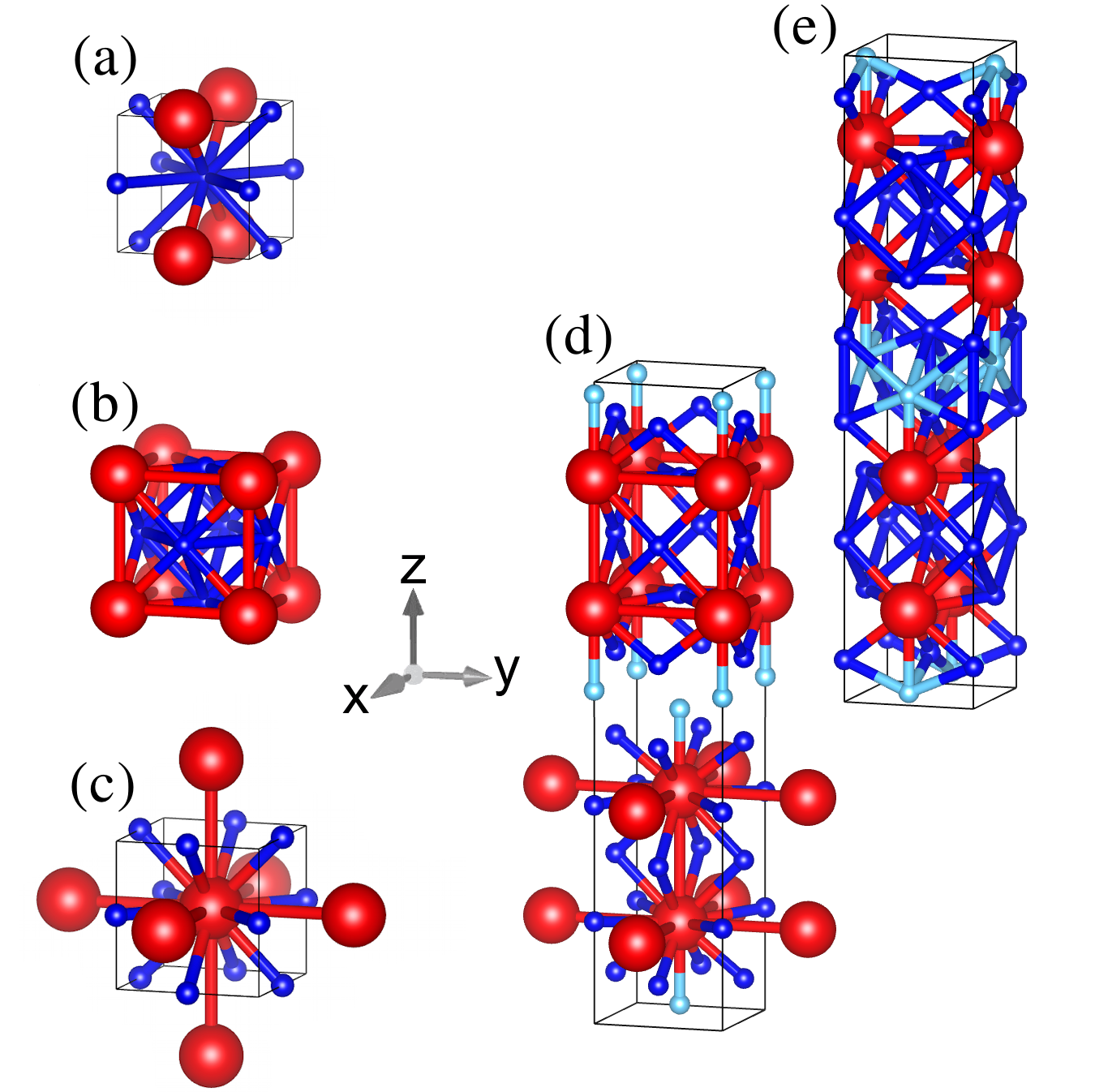}
\caption{FeH$_3$ (left) and FeH$_5$ (right) from different perspectives. Fe atoms are shown in red, H atoms are in blue and cyan, and nearest-neighbor bonds are shown as bicolor sticks. The FeH$_3$ cubic cell is chosen 
(a) with H in the middle and only its nearest-neighbor bonds shown,
(b) with an  fcc-like arrangement (Fe in the corners, H in the face centers) and all the nearest-neighbor bonds shown (Fe--Fe, H--H, and Fe--H), and  
(c) with Fe in the middle and the H--H nearest-neighbor bonds hidden, highlighting 12 H nearest neighbors within the cell and six Fe nearest neighbors in the six neighboring cells.
For FeH$_5$, we adopt a tetragonal unit cell, twice as large as the primitive cell, to help visual comparison to the ``parent'' structure FeH$_3$:  In cubic FeH$_3$, the Fe cages are stacked along each of the three $xyz$ directions; in FeH$_5$, instead of lining up along $z$, they are staggered to make room for the 13th (cyan) H atom, which saturates the vertical broken Fe--Fe bond of FeH$_3$. The two  visualizations of  FeH$_5$ correspond to (d) hiding the H--H bonds or (e) hiding the \mbox{Fe--Fe} bonds and horizontally shifting the origin along $x$ by half the horizontal lattice constant. 
}\label{fig:crystal}
\end{figure}

\section{Results and Discussion}
\begin{figure*}[t!]
	\includegraphics[width=2\columnwidth]{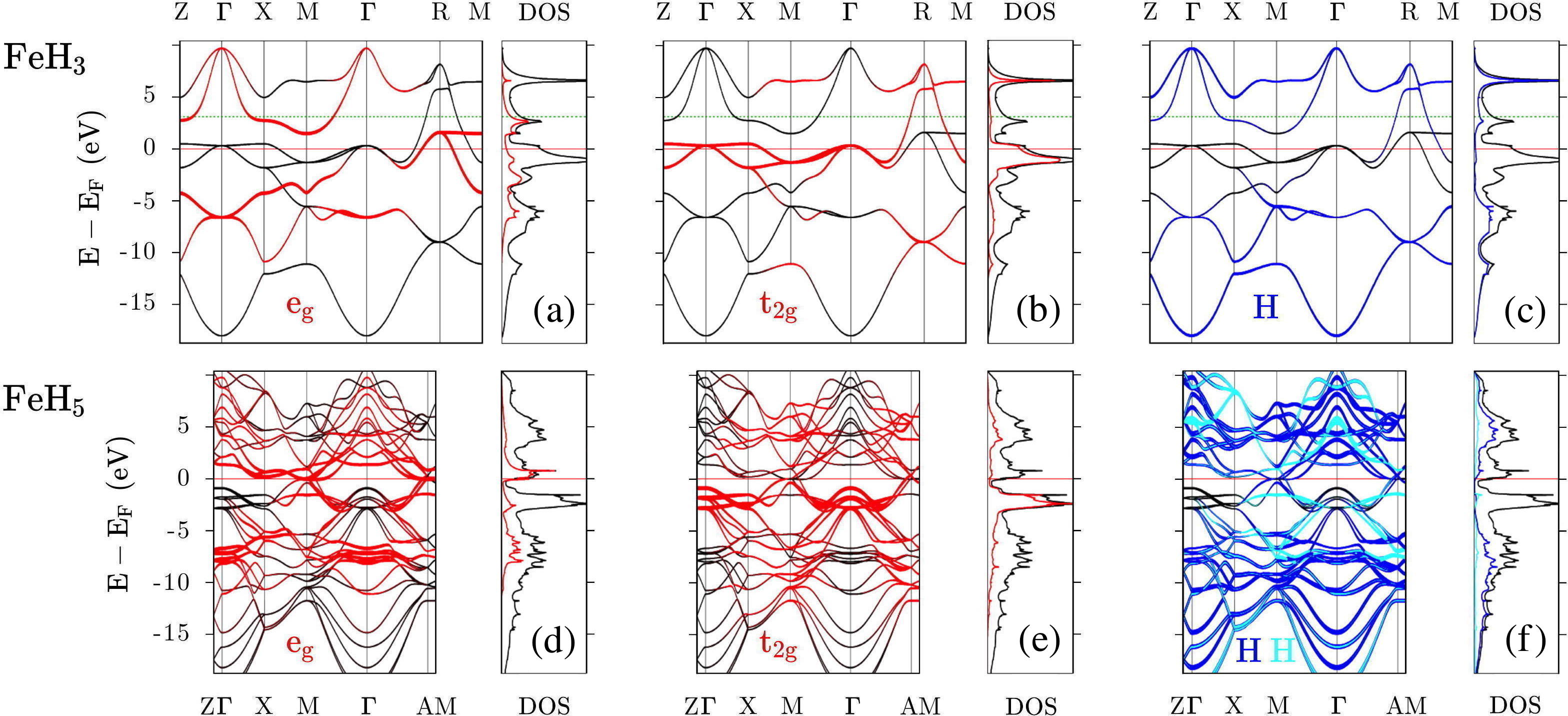}
	\caption{Electronic band structure and DOS of FeH$_3$ (top) and FeH$_5$ (bottom). The energy bands (black thin lines), decorated with dots whose size is proportional to their wave function character, are shown for (a) and (d) Fe~$e_g$, (b) and (e) Fe~$t_{2g}$, and (c) and (f) H~$1s$ states, along with the corresponding partial DOS. Total densities of states are shown in black, red indicates to Fe atoms, blue shows the 12 nearest-neighbor H of Fe which are present in both FeH$_3$ and FeH$_5$, and cyan indicates the 13th nearest neighbor of Fe which exists only in FeH$_5$ (see Fig.~\ref{fig:crystal}). The dashed green line indicates the energy needed to dope into bands with mixed ${\rm Fe}\hspace{0.05 cm}{e_g}\hspace{-0.02 cm}-\hspace{-0.02 cm}{\rm H}\hspace{0.05 cm}1s$ character. We use the notation for the special points of a simple tetragonal lattice for both FeH$_3$ and FeH$_5$ to facilitate comparison.}
	\label{fig:bands}
\end{figure*}

\begin{figure}
\includegraphics[width=1\columnwidth]{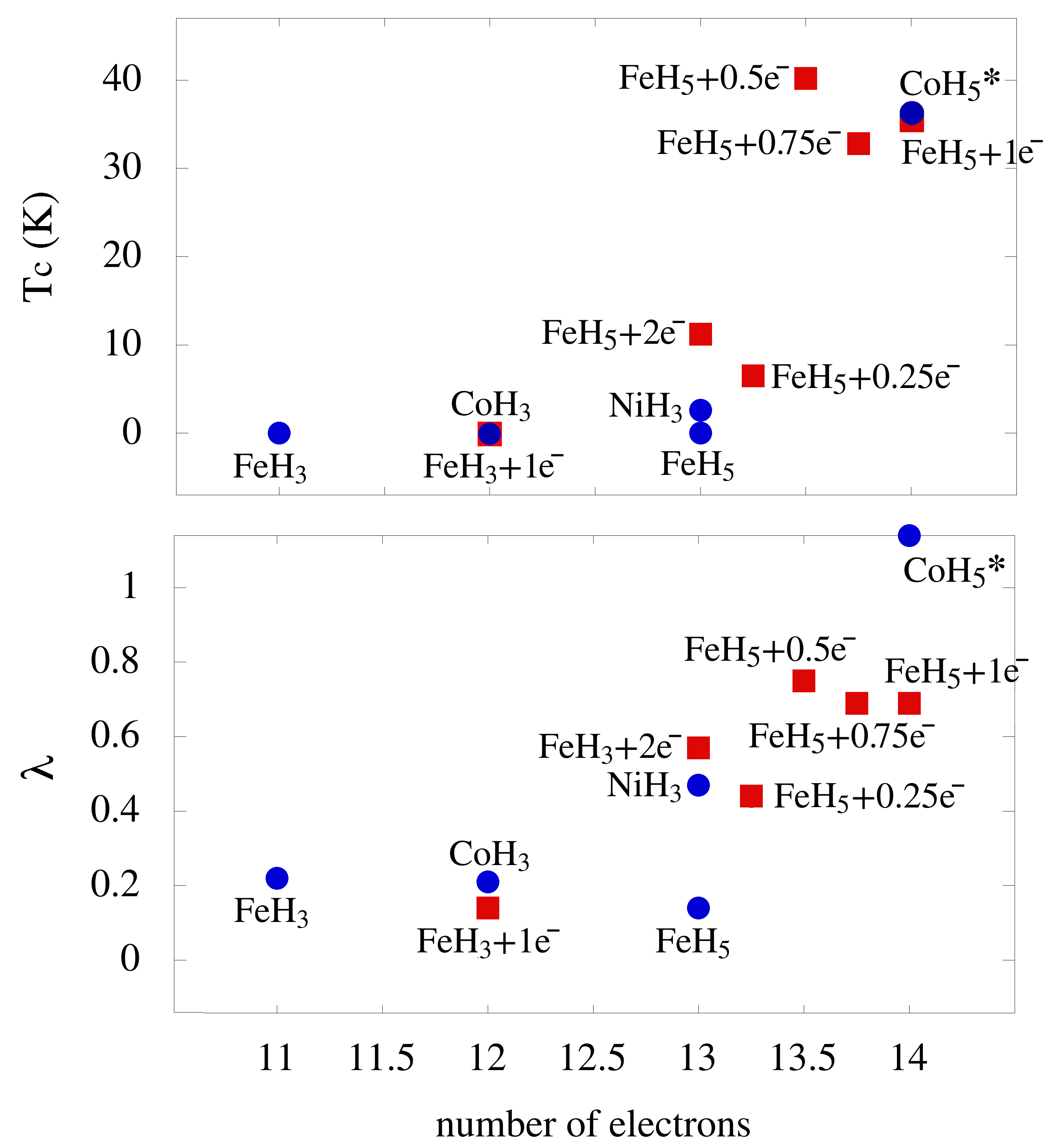}
	\caption{\label{fig:lambda_Tc} 
	Critical temperature $T_\text{c}$ (top panel) and \hbox{$e$-ph} coupling constant $\lambda$ (bottom panel) for a number of transition-metal polyhydrides. Blue dots indicate first principles results for stable FeH$_3$, CoH$_3$, NiH$_3$, and FeH$_5$ and for unstable CoH$_5^*$; red squares refer to rigid-band doping (see text).}
\end{figure}

According to P\'epin~\textit{et al.}~\cite{Pepin_FEH_2017}, FeH$_5$ forms above 130~GPa by hydrogenation of FeH$_3$, which is stable between 85 and 130~GPa. The structure was experimentally determined by measuring the x-ray diffraction patterns and performing a Rietveld refinement. Due to the low scattering power of the hydrogen atoms, only the Fe positions could be measured experimentally, and a structural search with density functional theory was employed by P\'epin~\textit{et al.} to find the lowest-enthalpy structure that is in agreement with both x-ray diffraction patterns and volume vs pressure curves from experiment~\cite{Pepin_FEH_2017}. Our own evolutionary structure searches with \textsc{uspex}~\cite{lyakhov2013new} confirm that the proposed structures for FeH$_3$ and FeH$_5$ are indeed stable in the considered pressure ranges~\cite{pepin_new_2014,Pepin_FEH_2017}. We depict the two crystal structures in Fig.~\ref{fig:crystal} and, in the following, will consider both compounds at the same pressure of 150~GPa. The computational details of all our calculations are listed in Appendix~\ref{app:comp_details}, convergence tests are provided in Appendix~\ref{app:conv_tests}, and the crystal structures for all considered compounds are given in Appendix~\ref{app:crystal_struct}.
\newline \indent
The left column of Fig.~\ref{fig:crystal} shows the crystal structure of FeH$_3$, which is \SC~\cite{pepin_new_2014}. In Fig.~\ref{fig:crystal}(c), where the \mbox{H--H} bonds are not shown, Fe (red) sits on a simple cubic lattice, surrounded by 12 H nearest neighbors (blue, bond length  \hbox{$\simeq$1.7 \AA})  and by the six nearest Fe atoms, a factor of $\sqrt{2}$ farther away than the H atoms yet closer than the bulk-iron Fe--Fe distance at normal pressure ({$\sim 2.5$~\AA}). Figure~\ref{fig:crystal}(a), where H is in the middle of the cube and the Fe--Fe bonds are hidden, highlights the local H environment, with eight H and four Fe equidistant nearest neighbors. Both Figs.~\ref{fig:crystal}(c) and \ref{fig:crystal}(a) are obvious consequences of the fcc-like structure of FeH$_3$ [Fig.~\ref{fig:crystal}(b)], i.e., a cubic cage with  Fe atoms in the corners and  H atoms in the face centers of the unit cell.
\newline \indent
The crystal structure of FeH$_5$, $I4/mmm$, determined by P\'epin \textit{et al.}~\cite{Pepin_FEH_2017} and confirmed by evolutionary crystal structure searches~\cite{Oganov_FEH}, is shown in the right-hand column of Fig.~\ref{fig:crystal}, where we adopt a 4-f.u. conventional unit cell, which is twice as large as the primitive unit cell. With this choice, it can be viewed as two cubic cages of Fe, cut out of bulk FeH$_3$ together with all their nearest-neighbor hydrogen atoms, vertically separated by a small void and displaced with respect to each other in the $xy$ plane. The empty space is occupied by additional hydrogen atoms, shown in cyan, which saturate the broken Fe--Fe bonds above and below. Thus, compared to the cubic FeH$_3$, FeH$_5$ appears as a stack of alternating layers of saturated  FeH$_3$-like cubic cages, where each Fe atom binds 13 hydrogen atoms (instead of 12) and five Fe atoms (instead of six). To emphasize this point of view, Fig.~\ref{fig:crystal}(d) shows the Fe--H and Fe--Fe bonds and hides the H--H bonds, thus emphasizing the similarities between FeH$_5$ and its precursor FeH$_3$. 
\newline \indent
Other authors argue that all of the hydrogen atoms (i.e., both blue and cyan), which lie in the void regions between subsequent layers of Fe cubic cages, should, instead, be regarded as a dense atomic-hydrogen layer~\cite{Pepin_FEH_2017,Majumdar_FEH}; this is not just a matter of taste,  as the two descriptions correspond to entirely different electronic (and hence superconducting) properties of the system.
We will show in the following that such a two-dimensional (2D) metallic-hydrogen scenario, in spite of its appealing implications for high-\tc \ conventional superconductivity~\cite{PhysRevLett.21.1748,PhysRevLett.92.187002}, is both geometrically and electronically unjustified.
\newline \indent
First of all, Fig.~\ref{fig:crystal}(e), by displaying a different set of nearest-neighbor bonds  with respect to Fig.~\ref{fig:crystal}(d) (i.e., by hiding the Fe--Fe bonds and displaying only the H--H and H--Fe bonds), reveals that the H network in the interstitial space between subsequent Fe cages (where the ``additional'' cyan hydrogen atoms also sit) has H--H distances ranging from 1.3 to 1.54~\AA, twice as large as the H$_2$ bond length (0.74~\AA) and also larger than the two \mbox{H--H} distances (0.98 and 1.2~\AA) predicted for solid atomic hydrogen at $\sim 500$~GPa~\cite{PhysRevB.93.174308}. In other words, the interstitial H network in FeH$_5$ is not much denser than the H network within the  iron cages, where H--H distances are all equal to {$\sim$1.6 \AA}  ({$\sim$1.65 \AA} in FeH$_3$).
\newline \indent
Second, all the Fe--H nearest-neighbor distances in FeH$_5$ (ranging from 1.46 to 1.70~\AA) are comparable to those found in FeH$_3$ (1.65~\AA) at the same pressure, the shortest one actually corresponding to the cyan interstitial H atom.
\newline \indent
Third, and most importantly, the 2D metallic-hydrogen scenario is not consistent with the electronic structure of FeH$_5$, where, on the one hand, no bands with hydrogen-only character may be found within at least $\pm 5$ eV from the Fermi level and, on the other hand, Brillouin-zone folding effects due to the 4-f.u. unit cell are enough to explain the main features of the FeH$_5$ band structure by tracing them back to the  simpler FeH$_3$ band structure. 
\newline \indent
This can be appreciated in Fig.~\ref{fig:bands}, where  the electronic band structures of FeH$_3$ (top) and FeH$_5$ (bottom), decorated with partial Fe and H characters, are shown.
\newline \indent
In FeH$_3$, the eight ${\rm Fe}\hspace{0.05 cm}3d\hspace{-0.02 cm}-\hspace{-0.02 cm}{\rm H}\hspace{0.05 cm}1s$ bands have a total bandwidth of $\sim 25$~eV. The
hydrogen $s$ states form one bonding band centered $\sim$ 15~eV below $E_F$ and two non bonding bands at higher energies [see Fig.~\ref{fig:bands}(c)]. The five Fe bands, subdivided into ${e_g}$ and ${t_{2g}}$ manifolds [see Figs.~\ref{fig:bands}(a) and \ref{fig:bands}(b), respectively], fall mostly within the wide ($\sim 10$~eV) gap between lower and upper H bands.
The Fermi level of FeH$_3$ cuts the band structure in the middle of the ${t_{2g}}$ manifold, where the hybridization of ${\rm H}\hspace{0.05 cm}1s$  with ${\rm Fe}\hspace{0.05 cm}3d$ states is negligible, with negligible H contribution to the density of states (DOS). The hybridization with hydrogen $1s$ states is, instead, significant in the ${e_g}$ manifold, located 2.5~eV above and 7.5~eV below the Fermi level.
\newline \indent
In FeH$_5$, the projection on Wannier orbitals shows that the electronic bands around the Fermi level are mainly of ${\rm Fe}\hspace{0.05 cm}3d$ character in  either its ${e_g}$ or ${t_{2g}}$ representation [see Figs.~\ref{fig:bands}(d) and \ref{fig:bands}(e), respectively]. Compared to FeH$_3$, there are two more electrons per f.u. in FeH$_5$, and hence the Fermi level cuts the band structure at the top of the ${t_{2g}}$ manifold, in a region where the electronic DOS is extremely low and exhibits a pseudogap. Such a Fermi-level shift due to two more electrons per f.u. leaves, however, the contribution of ${\rm H}\hspace{0.05 cm}1s$ states to the DOS [see Fig.~\ref{fig:bands}(f)] at the Fermi level as low as in FeH$_3$.
As a result, for both FeH$_3$ and FeH$_5$, the dominant contribution to the DOS at the Fermi level, and thus to superconducting pairing, comes from the Fe sublattice, not from the H sublattice. This in turn implies that, in the best case, these iron hydrides will behave like elemental metals (\tc \ $ \lesssim 10$~K) and not like the recently discovered high-pressure superconducting hydrides (\tc \ $ \gtrsim 77$~K)~\cite{DrozdovEremets_Nature2015, Flores-Livas2016,Duan_SciRep2014,PhysRevLett.114.157004,
Drozdov_PH3_arxiv2015,Flores_PH3_PRBR2016,shamp_decomposition_2015,Fu_Ma_pnictogenH_2016}.
\begin{figure}
	\includegraphics[width=1\columnwidth,angle=0]{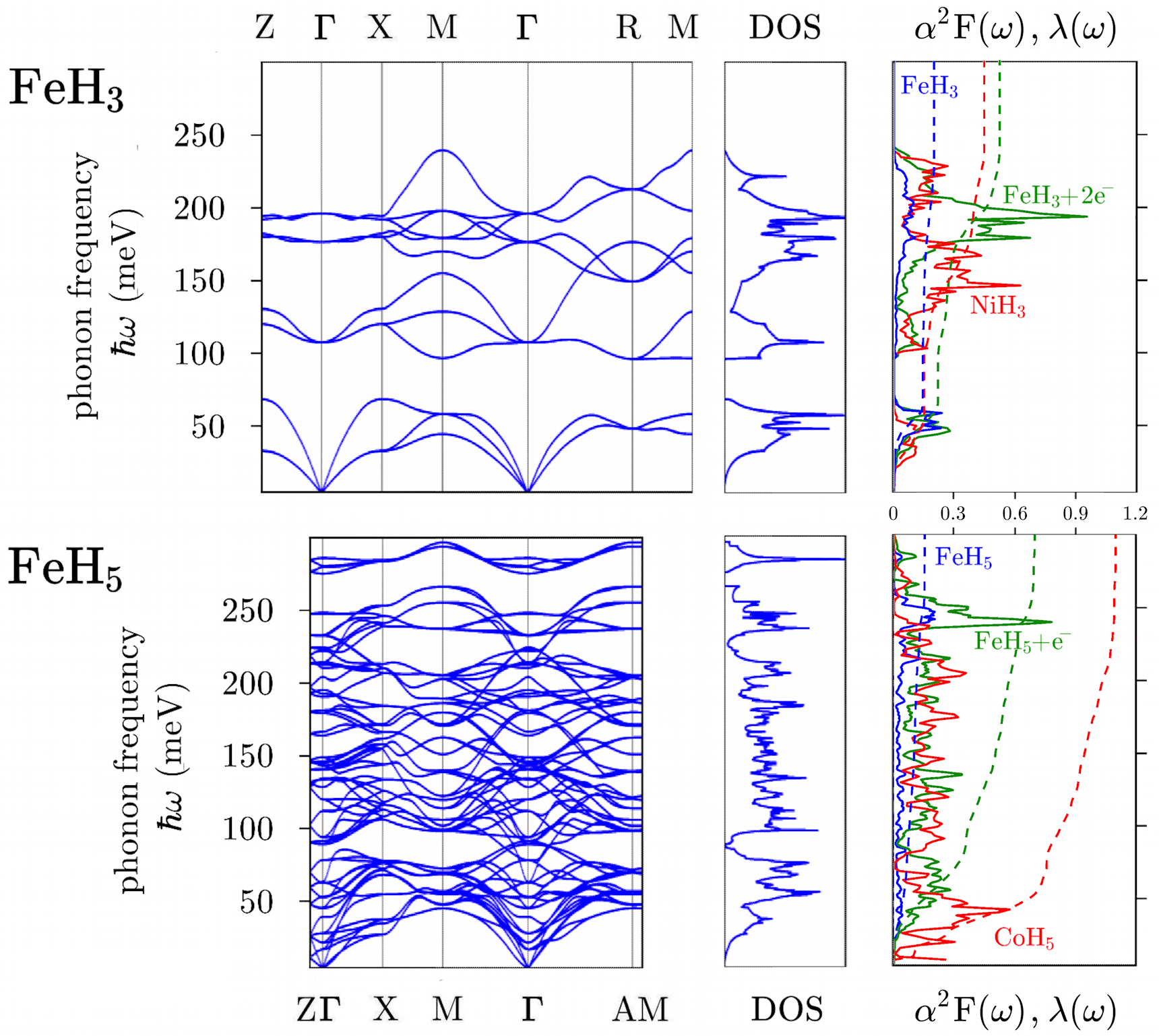}
	\caption{\label{fig:phonons} Phonon dispersion (left), phonon DOS (middle), and Eliashberg function $\alpha^2 F(\omega)$ [right, Eq.~(\ref{eq:alpha})]. The top panels refer to FeH$_3$ (blue curves), where Eq.~(\ref{eq:alpha}) was also evaluated for two-electron-doped FeH$_3$ (green) and for NiH$_3$ (red).
		The dashed curves indicate the frequency-dependent coupling constant $\lambda(\omega)$. The bottom panels refer to FeH$_5$ (blue curves), where Eq.~(\ref{eq:alpha}) was also evaluated for one-electron-doped FeH$_5$ (green) and unstable CoH$_5$ (red; see text).}
\end{figure}
\begin{figure*}[t]
	\includegraphics[width=2.0\columnwidth]{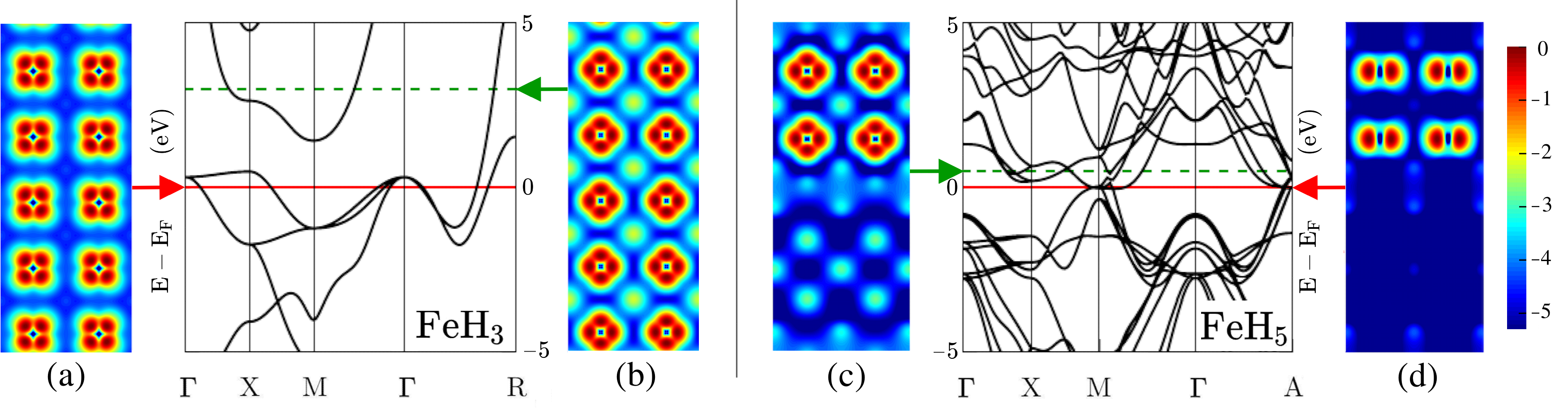}
	\caption{Low-energy band structure and LDOS [Eq.~(\ref{eq:LDOS})]  
		for (a) and (b) FeH$_3$ and (c) and (d) FeH$_5$ along a (100) plane, which cuts
		through Fe and H atoms in both compounds (see Fig.~\ref{fig:crystal}).
		In (a) and (d) the LDOS is displayed at the Fermi level of the corresponding compounds (red arrows); in (b) and (c) the LDOS is displayed at the energy of an empty Fe--H hybrid band above the respective Fermi levels (green arrows; see text), corresponding to an electron count of FeH$_3 + 2 e^-$ and FeH$_5 + 0.5 e^-$ in Fig.~\ref{fig:lambda_Tc}. The color scale used corresponds to $\ln{[\text{LDOS}/\text{max}(\text{LDOS})]}$ to improve visibility.
		\label{fig:LDOS}}
\end{figure*}
\newline \indent
The above qualitative prediction is quantitatively confirmed by our {\em ab initio} Migdal-Eliashberg calculations. Figure~\ref{fig:phonons} shows, for FeH$_3$ (top) and FeH$_5$ (bottom), the phonon dispersion, the phonon DOS, and the Eliashberg $e$-ph spectral function, from whose moments we estimate a superconducting $T_\text{c}$ using the McMillan-Allen-Dynes formula~\cite{allen_transition_1975}:
\begin{equation}
\label{eq:McMillan}
  T_\text{c}=\frac{\omega_\text{log}}{1.2 k_B}\exp\left[-\frac{1.04(1+\lambda)}{\lambda-\mu^{*}(1+0.62\lambda)}\right]~,
\end{equation}
where $\omega_\text{log}$ and $\lambda$ are the logarithmic-averaged phonon frequency and the $e$-ph coupling constant, respectively, and $\mu^*$ is the Coulomb (Morel-Anderson) pseudopotential. Setting $\mu^*$ to a typical value ($\mu^*=0.16$), we obtain $\omega_\text{log}=65.7$~meV, $\lambda=0.2$, $T_\text{c}=0$~K for FeH$_3$ and $\omega_\text{log}=90.5$~meV, $\lambda=0.14$, $T_\text{c}=0$~K for FeH$_5$. In other words, for both compounds the isotropic version of the Migdal-Eliashberg theory predicts no conventional superconductivity (we double-checked that this result holds even within the fully anisotropic theory)~\cite{migdal_interaction_1958,eliashberg_interactions_1960,ponce_epw:_2016}.
\newline \indent
Our results for FeH$_3$ agree with those of other authors~\cite{Oganov_FEH}.
For FeH$_5$, instead, our findings are in remarkable disagreement with two previous studies on the same compound, which both predict a substantial \tc \ of around 50~K~\cite{Majumdar_FEH,Oganov_FEH}. As detailed in Appendix~\ref{app:conv_tests}, we have tested several possible sources of discrepancy, but all calculations with physically justifiable parameters invariably yielded a vanishing \tc \ for FeH$_5$. Our study on the dependence of \tc \ on doping in the rigid-band approximation corroborates these findings. 
\newline \indent
Before presenting this additional study, we want to explain how the superconducting
trends in FeH$_3$ and FeH$_5$ can be understood on  {\em qualitative} grounds.
The main ingredient of the conventional theory of superconductivity is the Eliashberg function
\begin{equation}
 \alpha^2 F(\omega) = \frac{1}{N(E_F)} \sum \limits_{\mathbf{k} \mathbf{q},\nu} 
|g_{\mathbf{k},\mathbf{k}+\mathbf{q},\nu}|^2 \delta(\epsilon_\mathbf{k}) 
\delta(\epsilon_{\mathbf{k}+\mathbf{q}}) \delta(\omega-\omega_{\mathbf{q},\nu})~, 
\label{eq:alpha}
\end{equation}
from which the parameters of the McMillan-Allen-Dynes formula~(\ref{eq:McMillan}) are obtained as: $\lambda=2 \int \frac{d\omega}{\omega} {\alpha^2 F(\omega)}$, $\omega_\text{log}=\exp\left[\frac{2}{\lambda}\int \frac{d\omega}{\omega} \alpha^2 F(\omega) \ln(\omega) \right]$. In Eq.~(\ref{eq:alpha}), $N(E_F)$ is the DOS at the Fermi level, $\omega_{\mathbf{q},\nu}$ is the phonon frequency of mode $\nu$ and wave vector $\mathbf{q}$, and $|g_{\mathbf{k},\mathbf{k}+\mathbf{q},\nu}|$ is the $e$-ph matrix element between two electronic states of wave vectors $\mathbf{k}$ and $\mathbf{k+q}$ at the Fermi level~\cite{AllenMitrovic1983}.
The double-delta function $\delta (\varepsilon_{\mathbf{k}}^n)\delta (\varepsilon_{\mathbf{k+q}}^m)$ restricts the sum of $e$-ph matrix elements to electronic states at the Fermi level.
\newline \indent
High-\tc \ conventional superconductors are compounds where the double-$\delta$ function in Eq.~(\ref{eq:alpha}) selects electronic states with a large $|g_{\mathbf{k},\mathbf{k}+\mathbf{q},\nu}|$, i.e., electronic states which are strongly modified by the ionic motion. A real-space-resolved electronic DOS, the so-called local density of states (LDOS), defined as
\begin{equation}
N(E,{\bf r})=\frac{1}{(2 \pi)^3}\sum_n \int d^3k \delta (E - \varepsilon^n_{\bf k})| \psi_{n {\bf k}}({\bf r})|^2
\label{eq:LDOS}
\end{equation}
and evaluated at the Fermi level, provides visual intuition of why the $e$-ph coupling is large or small in a given compound. Also, when evaluated at other selected energies, it may tell something about the $e$-ph coupling of a particular band (or band manifold).
To this end we present in Fig.~\ref{fig:LDOS},
for FeH$_3$ [Fig.~\ref{fig:LDOS}(a)] and FeH$_5$ [Fig.~\ref{fig:LDOS}(d)],
the LDOS at the Fermi level $N(E_F,{\bf r})$ along the (100) lattice plane, which cuts through Fe and H atoms in both FeH$_3$ and FeH$_5$ [see  Figs.~\ref{fig:crystal}(b) and \ref{fig:crystal}(d)].
A red arrow connects each of these two panels with the red horizontal line highlighting the Fermi level of the corresponding band structure.
\newline \indent
Since, as already mentioned when discussing Fig.~\ref{fig:bands}, the Fermi level falls within the Fe ${t_{2g}}$ bands (in FeH$_3$ among them, in FeH$_5$ at their top) and since the H contribution to the electronic DOS at $E_F$ is negligible in both compounds, it is readily understood that the LDOS pattern, although qualitatively different (since $E_F$ does not cut the $t_{2g}$ bands in the same place), appears intense around the Fe atoms in both compounds, and either undetectable [FeH$_3$, Fig.~\ref{fig:LDOS}(a)] or barely visible [FeH$_5$, Fig.~\ref{fig:LDOS}(d)] around the H atoms and in the interstitial regions. From these Fe-dominated bands we expect almost no coupling to the H motion and an $e$-ph coupling to the Fe motion as low as in bulk iron (not a superconductor). Indeed, in both in the top (FeH$_3$) and bottom (FeH$_5$) panels of Fig.~\ref{fig:phonons}, the corresponding Eliashberg functions $\alpha^2 F(\omega)$ show a very low average $e$-ph coupling, uniformly spread over Fe and H modes.
\newline \indent
Not all of the electronic states in these transition-metal polyhydrides, however, have such a poor intrinsic $e$-ph coupling.
If, by doping, one added more electrons to the system, one could completely fill the ${t_{2g}}$ bands, and the Fermi level would eventually reach a band which has a mixed ${\rm Fe}\hspace{0.05 cm}{e_g}\hspace{-0.02 cm}-\hspace{-0.02 cm}{\rm H}\hspace{0.05 cm}1s$ character. In FeH$_3$ such an \mbox{Fe--H} hybrid band, highlighted by a horizontal dashed green line in Fig.~\ref{fig:LDOS} (and also in Fig.~\ref{fig:bands}), starts at $\sim 2$~eV above the Fermi level along the $X$--$M$ segments. In FeH$_5$, due to the different electron count and to Brillouin-zone folding effects, the same Fe--H hybrid band can be found just $\sim 0.5$ eV above $E_F$ (other replicas appear above it as well). 
\newline \indent
In Fig.~\ref{fig:LDOS}, a green arrow connects the mean energy of each of these bands to the corresponding plot of the LDOS in the (100) lattice plane, shown in the Figs.~\ref{fig:LDOS}(b) and \ref{fig:LDOS}(c) for FeH$_3$ and FeH$_5$, respectively. For both compounds, the LDOS displays the same (${e_g}$) symmetry around the Fe atom at this energy, and unlike the bands at the Fermi level, it also displays a considerable weight (bright yellow spots) around the H atoms.
For electronic states with such a real-space distribution, it is reasonable to expect a sizable $e$-ph coupling since they are likely to be affected by both H and Fe vibrations. The real-space distribution itself suggests, however, that
the $e$-ph coupling  will be moderate because of the lack of directional
bonds between Fe and H. As a matter of fact, the LDOS at the Fermi level in Fig.~\ref{fig:LDOS} shows no evidence of any directional bonds at all, as if all sticks between pairs of atoms had been removed in Fig.~\ref{fig:crystal}.
\newline \indent
A rough quantitative estimate of the $e$-ph coupling of the aforementioned  Fe--H hybrid electronic states may be  given by simply recomputing the Eliashberg function~(\ref{eq:alpha}) after a rigid shift of the Fermi level into that energy band. Figure~\ref{fig:lambda_Tc} shows as red squares the calculated $e$-ph and superconducting properties (i.e., $\lambda$ and $T_\text{c}$) as a function of electron count for our two compounds. We see that the threshold to bring the Fermi level into the Fe--H hybrid band is between 13 and 14 $e^{-}$/f.u., which implies at least two additional $e^{-}$/f.u. in FeH$_3$, but only $\sim$0.25 $e^{-}$/f.u. in FeH$_5$. We also see that, in agreement with our expectations, the $e$-ph coupling experiences an abrupt increase as soon as the respective doping thresholds are exceeded.
\newline \indent
This numerical rigid-band experiment is instructive for understanding the key features of the FeH$_3$/FeH$_5$ electronic structure and $e$-ph coupling but is not a practical way to improve the superconducting properties of real materials. This  
can, instead, be achieved by {\em chemical} means, i.e., by replacing Fe with other transition elements with more $d$ electrons.
To this end, we have studied the $e$-ph coupling properties of $X$H$_3$ compounds in the \SC \ structure, with $X$ = Co ($+1 e^-$), Ni ($+2 e^-$), and Cu ($+3 e^-$).
For $X$ = Co and Ni we obtain a dynamically stable \SC \ structure, and evolutionary structure searches consistently converge to such a \SC \ structure at 150~GPa. Cu, on the other hand, is dynamically unstable, and in this case analogous searches run into highly distorted structures. 
\newline \indent
The results of these simulations are shown as blue dots in Fig.~\ref{fig:lambda_Tc}, where our previous rigid-band-doping estimates are marked as red squares. Our results for actual compounds follow the rigid-band-doping trends reasonably, confirming our understanding of the $e$-ph mechanism in this class of compounds. In particular, we see that the addition of electrons to  FeH$_3$ (by replacing Fe with Ni) is a more effective way to increase $\lambda$ (bottom panel) than the addition of H atoms, as in FeH$_5$. 
In connection with these calculations, we also report the  Eliashberg functions for FeH$_3$, FeH$_3$ + $2e^{-}$, and NiH$_3$ in the top right panel of Fig.~\ref{fig:phonons}, which corroborates our findings: Moving the Fermi level more and more into the Fe--H hybrid band causes a progressive increase of the $e$-ph coupling at high energies, where Fe--H modes are concentrated.
Indeed, as soon as two electrons are added to the system ( FeH$_3$ + $2e^{-}$, green) a large peak centered at $\sim 180$ meV appears, which is absent in
FeH$_3$ (blue). This peak shifts to lower energies ($\sim 150$ meV) in NiH$_3$ (red) because the corresponding phonon frequencies are shifted down by electronic screening effects (see Appendix~\ref{app:NiH3_CoH5}).
In the top panel of Fig.~\ref{fig:lambda_Tc} we see, however, that
the actual substitution of Fe with Ni is not sufficient to yield appreciable superconductivity. In spite of the larger $\lambda$ in NiH$_3$, the renormalization of phonon frequencies and matrix elements due to the electronic
screening brings $T_\text{c}$ down to $\sim$ 3~K, compared to the 11~K of the
corresponding rigid-band result.
Adding more electrons amplifies such effects, making CuH$_3$ (not shown in Fig.~\ref{fig:lambda_Tc}) dynamically unstable in this structure, as mentioned previously.
\newline \indent
For similar reasons CoH$_5$, which, according to our rigid-band prediction, should be an $\sim 35$ K superconductor (in the same $I4/mmm$ structure as FeH$_5$), turns out to be dynamically unstable in the harmonic approximation, with imaginary phonon frequencies over large portions of the Brillouin zone (see Appendix~\ref{app:NiH3_CoH5}), indicating that CoH$_5$ will, most likely, \emph{not} form in this structure at this pressure. Nevertheless, if we compute its $e$-ph coupling properties by integrating the Eliashberg function over the real portion of the phonon spectrum, we obtain a large value for the coupling $\lambda$ and a \tc \ considerably larger than that of FeH$_5$. The bottom panels of Fig.~\ref{fig:phonons} show a comparison of $\alpha^2 F(\omega)$ of CoH$_5$ with the corresponding rigid-band result; here, too, we observe increased coupling at large frequencies, but in CoH$_5$ part of the corresponding spectrum is shifted to negative $\omega^2$ (imaginary frequencies), which we exclude from the integral. This explains why the total $e$-ph coupling of CoH$_5$ is lower than that of the rigid-band calculation. We did not pursue the CoH$_5$ experiment any further for two reasons: ($i$) the presence of the Fe--H hybrid band depends crucially on the crystal structure, and the dynamical instability we found indicates an important lattice distortion. ($ii$)
Even if, in principle, anharmonic effects, which can be sizable in high-pressure hydrides~\cite{errea2016quantum}, could stabilize the $I4/mmm$ structure, in practice, a recent study~\cite{wang2017high} found that the CoH$_5$ stoichiometry is not thermodynamically stable up to 300~GPa; that is, it would be extremely difficult to stabilize it in other structures as well. What we found so far is enough to strongly suggest that a sizable \tc \ cannot be obtained by simple chemical means in $X$H$_5$ compounds.

\section{Conclusions}
In summary, we have computed from first principles the electronic, structural, and superconducting properties of the new iron hydride FeH$_5$, recently synthesized by P\'epin \textit{et al.}~\cite{Pepin_FEH_2017} At variance with two previous studies~\cite{Majumdar_FEH, Oganov_FEH}, we found that FeH$_5$ is not superconducting and showed that this had to be expected since its electronic states at the Fermi level, dominated by $d$--metal bonding, have an intrinsically low $e$-ph coupling. Moreover, the only band with appreciable Fe--H hybridization and $e$-ph coupling lies above the Fermi level and is thus inaccessible to superconductivity. We exploited doping to shift the Fermi level into this Fe--H hybrid band and, indeed, obtained a higher \tc \ but found that, even so, {\tc} would hardly reach 40~K. Moreover, even this value is practically impossible to achieve, as the required doping levels push these compounds beyond their structural stability limits. The picture emerging from our analysis strongly contradicts the notion, proposed by other authors~\cite{Pepin_FEH_2017,Majumdar_FEH}, of a layer of dense metallic hydrogen dominating the superconducting properties of FeH$_5$.
Our results do not translate into the identification of new candidates for high-\tc \ superconductivity in the FeH$_x$ family; on the contrary, they rule out this class of compounds from the list of potential high-\tc, high-pressure superconductors,  shedding new light on the mechanisms leading to high \tc \ in high-pressure hydrides.

\section{acknowledgments}
This work was supported by Austrian Science Fund (FWF) Projects No. J 3806-N36 and P 30269-N36, the dCluster of the Graz University of Technology, and the VSC3 of the Vienna University of Technology. L.B. and G.B.B. acknowledge support from Fondo Ateneo-Sapienza 2017.
%\end{acknowledgments}

\appendix

%\vfill

%\clearpage

\section{Computational details}
\label{app:comp_details} 
Our calculations were carried out using optimized norm-conserving Vanderbilt pseudopotentials~\cite{hamann_optimized_2013,schlipf_optimization_2015} within the Perdew-Burke-Ernzerhof functional~\cite{perdew_generalized_1996} that include the semicore electrons of Fe.
We employed the \textsc{quantum espresso} package~\cite{giannozzi_quantum_2009} for the electronic structure and lattice dynamics, the \textsc{epw} code~\cite{ponce_epw:_2016} for the $e$-ph interaction and the superconducting properties, and the \textsc{wannier90} code~\cite{mostofi_wannier90_2008} for generating maximally localized Wannier functions. The vibrational properties were obtained using density functional perturbation theory. In all calculations for the density functional theory ground state we used a kinetic cutoff for the plane waves of 65~Ry and a Gaussian smearing $\sigma$ of 0.01~Ry. For FeH$_3$, we sampled the Brillouin zone for the electronic properties using a $24 \times 24 \times 24$ grid and an $8 \times 8 \times 8$ grid for the vibrational properties. Within \textsc{epw}, all quantities were interpolated onto $30 \times 30 \times 30$ grids using eight Wannier functions. For FeH$_5$ in the 4-f.u. cell (2-f.u. cell), we sampled the Brillouin zone for the electronic properties using a $24 \times 24 \times 6$ ($24 \times 24 \times 12$) grid and an $8 \times 8 \times 2$ ($6 \times 6 \times 3$) for the vibrational properties. Within \textsc{epw}, all quantities were interpolated onto $32 \times 32 \times 8$ ($30 \times 30 \times 10$) grids using 40 (20) Wannier functions. In all calculations for superconducting properties in \textsc{epw}, the Matsubara frequency cutoff was set to 1~eV, and the Dirac $\delta$ were replaced by Lorentzians with a width of 25~meV (electrons) and 0.05~meV (phonons).

\section{Convergence tests}
\label{app:conv_tests}

\begin{figure}[t]
	\includegraphics[width=1\columnwidth]{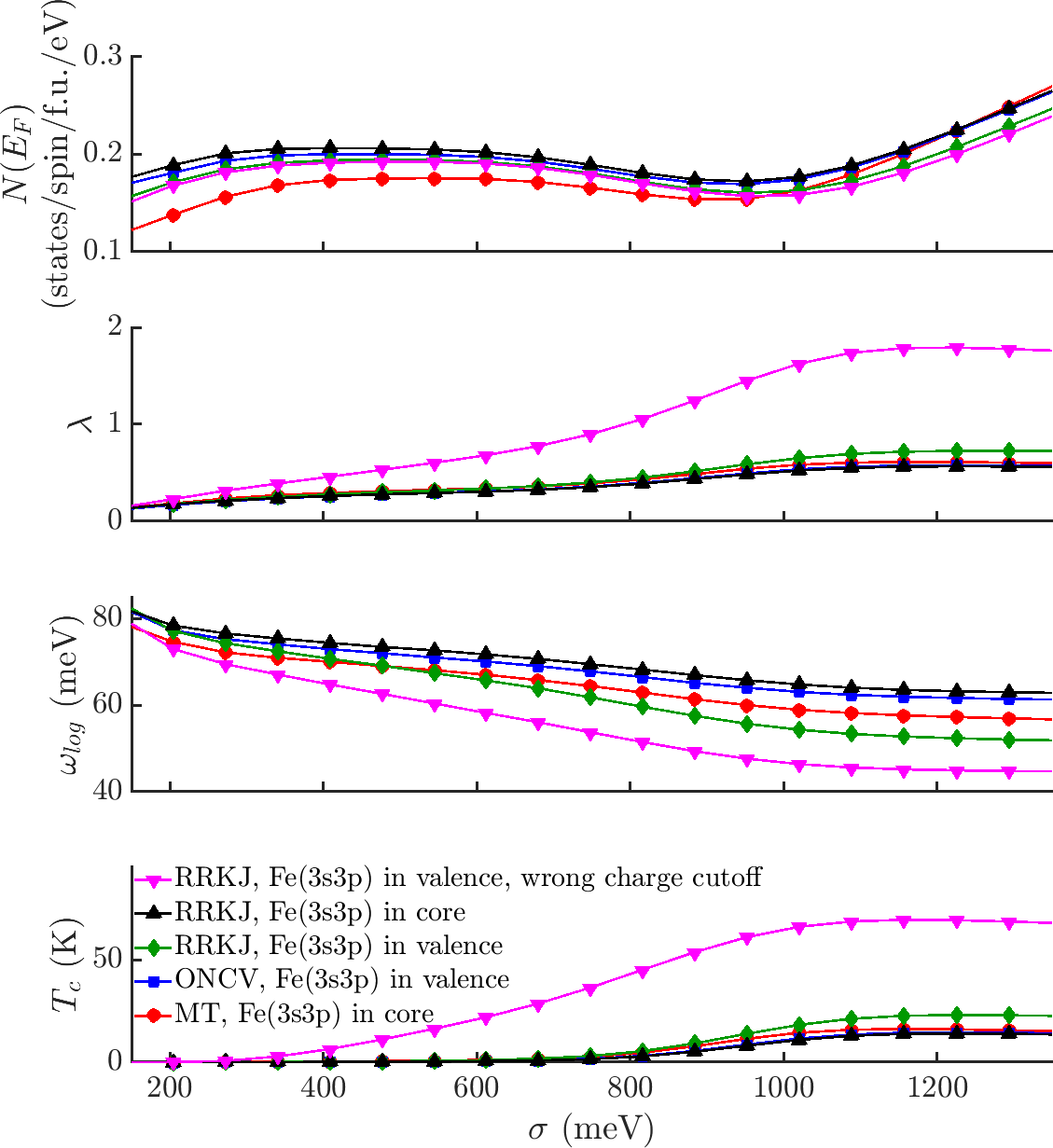}
	\caption{$N(E_F)$ (first panel), $\lambda$ (second panel), $\omega_\text{log}$ (third panel) and $T_\text{c}$ (fourth panel) of FeH$_5$ as a function of electronic smearing $\sigma$ and the pseudopotential~\cite{hamann_optimized_2013,rappe1990optimized,troullier1991efficient}.} \label{fig:S1}
\end{figure}

As our results for the superconducting parameters of FeH$_5$ are in variance with Refs.~\cite{Oganov_FEH,Majumdar_FEH}, we performed several tests to check our results. First, since the data presented in the main text were obtained with the \textsc{epw} code~\cite{ponce_epw:_2016}, we checked that a calculation strictly within the \textsc{quantum espresso} package~\cite{giannozzi_quantum_2009}, as performed in Refs.~\cite{Oganov_FEH,Majumdar_FEH}, leads to the same results. In this case, we get $\lambda = 0.22$, $\omega_\text{log} = 87.6$~meV, and $T_\text{c} = 0$~K, in good agreement with the results from the more elaborate \textsc{epw} calculation ($\lambda=0.14$, $\omega_{\log}=90.5$~meV,  $T_c=0$~K). The difference between the results can be explained by taking into account that in \textsc{epw}, we used much denser Brillouin-zone grids for the integration of both electronic and vibrational properties, which allowed us to use much smaller smearing parameters (25~meV for electrons and 0.05~meV for phonons). Having established the reliability of the \textsc{quantum espresso} results, all further convergence tests are performed using only this code, as employed in Refs.~\cite{Oganov_FEH,Majumdar_FEH}.
\newline \indent
In Fig.~\ref{fig:S1}, we show the DOS at the Fermi level $N(E_F)$ (first panel), electron-phonon coupling $\lambda$ (second panel), logarithmic-averaged phonon frequency $\omega_\text{log}$ (third panel), and superconducting critical temperature $T_\text{c}$ (fourth panel) of FeH$_5$ as a function of electronic smearing $\sigma$ and pseudopotential (PP). 
The tested PPs include the norm-conserving Martins-Troullier~\cite{troullier1991efficient} (MT) type, the optimized norm-conserving Vanderbilt~\cite{hamann_optimized_2013} (ONCV) type, and the ultrasoft Rappe-Rabe-Kaxiras-Joannopoulos~\cite{rappe1990optimized} (RRKJ) type.
For these convergence calculations, we used the \mbox{4-f.u.} cell, a Brillouin-zone grid of $16 \times 16 \times 4$ for the electronic properties, and a $4 \times 4 \times 1$ grid for the vibrational properties to match the grids of Refs.~\cite{Oganov_FEH,Majumdar_FEH}. We used a kinetic-energy cutoff of $65$~Ry for the plane waves and $\mu^* = 0.1$ to calculate $T_\text{c}$.
\begin{table}[t]
	\caption{\label{tab:crystal_structures} Crystal structures of the compounds mentioned in the main text. The atomic positions are given in crystal coordinates, along with their Wyckoff positions in square brackets.}
	\begin{center}
		\resizebox{\columnwidth}{!}{%
			\begin{tabular}{c c c c c c }
				\hline \hline
				& Space & Lattice          &      &  & Atomic positions \\
				& group & parameters (\AA) & Atom & Wyckoff & Crystal  \\ \hline
				\multirow{ 2}{*}{FeH$_3$} & \multirow{ 2}{*}{\textit{$Pm$\={3}$m$}} & \multirow{ 2}{*}{$a=b=c=2.33$} & Fe & [1a] &  (0.00,  0.00,  0.00) \\
				& 										&                                & H  & [3c] &  (0.00,  0.50,  0.50) \\ 
				\multirow{ 2}{*}{CoH$_3$} & \multirow{ 2}{*}{\textit{$Pm$\={3}$m$}} & \multirow{ 2}{*}{$a=b=c=2.33$} & Co & [1a] &  (0.00,  0.00,  0.00) \\
				&                                         &                                & H  & [3c] &  (0.00,  0.50,  0.50) \\ 
				\multirow{ 2}{*}{NiH$_3$} & \multirow{ 2}{*}{\textit{$Pm$\={3}$m$}} & \multirow{ 2}{*}{$a=b=c=2.35$} & Ni & [1a] &  (0.00,  0.00,  0.00) \\
				&                                         &                                & H  & [3c] &  (0.00,  0.50,  0.50) \\ 
				\multirow{ 5}{*}{FeH$_5$} & \multirow{ 5}{*}{\textit{$I4/mmm$}}     &                                & Fe & [4e] &  (0.00,  0.00,  0.10) \\
				&                                         &   $a=b=2.39$                   & H  & [8g] &  (0.00,  0.50,  0.32) \\
				&                                         & \multirow{ 1}{*}{$c=11.50$}    & H  & [4e] &  (0.00,  0.00,  0.59) \\
				&                                         &                                & H  & [4c] &  (0.00,  0.50,  0.00) \\
				&                                         &                                & H  & [4e] &  (0.00,  0.00, 0.77) \\ 
				\multirow{ 5}{*}{CoH$_5$} & \multirow{ 5}{*}{\textit{$I4/mmm$}}     &                                & Co & [4e] &  (0.00,  0.00,  0.10) \\
				&                                         &   $a=b=2.40$                   & H  & [8g] &  (0.00,  0.50,  0.32) \\
				&                                         & \multirow{ 1}{*}{$c=11.39$}    & H  & [4e] &  (0.00,  0.00,  0.59) \\
				&                                         &                                & H  & [4c] &  (0.00,  0.50,  0.00) \\
				&                                         &                                & H  & [4e] &  (0.00,  0.00, 0.77) \\ \hline \hline
			\end{tabular}
		}
	\end{center}
\end{table}
\newline \indent
In the case of the norm-conserving PPs, the MT type (red lines with circles) considers the Fe $3s$ and $3p$ states to be in the core, while the ONCV type (blue lines with squares) includes these states in the valence. Despite the differences, however, both these PPs lead to very similar results with respect to the superconducting properties, as is apparent from Fig.~\ref{fig:S1}. The same holds true when comparing the results for the ultrasoft RRKJ with the Fe $3s$ and $3p$ states in the core and in the valence (black line with upwards-pointing triangles and green line with diamonds, respectively), where only little differences are appreciable.
As a function of electronic smearing $\sigma$, we find that all previously discussed PPs yield very similar results: For $\sigma < 500$~meV we get an $e$-ph coupling $\lambda < 0.3$ and values for $T_\text{c} < 1$~K. Even for extremely large $\sigma > 1000$~meV, which is definitely too large to obtain reasonable results, the values for $\lambda$ are below 0.75 and for $T_\text{c} < 23$~K. We also find that the effects of using a larger $8 \times 8 \times 1$ Brillouin-zone grid for the vibrational properties on the superconducting properties are small. For example, we find for the ONCV case that for a smearing of 270~meV, $\lambda$ increases from 0.21 to 0.26, and $T_\text{c}$ increases from 0 to 1~K. These differences, which keep decreasing with increasing $\sigma$, can be considered negligible for our current discussion.

However, when using ultrasoft pseudopotentials, one needs to choose a considerably larger energy cutoff for the electron density than with norm-conserving PPs, usually in the range of around 12 times the kinetic-energy cutoff. The previously discussed green and black curves, for example, were calculated using a density cutoff of 800~meV. The effect of using a too small electron density cutoff can be seen in the magenta line with downwards-pointing triangles in Fig.~\ref{fig:S1}, where we used a too small electron density cutoff of only 260~meV, i.e., only 4 times the kinetic-energy cutoff. Only in this case and using very large values for $\sigma$ were we able to reproduce the results of Refs.~\cite{Oganov_FEH,Majumdar_FEH}, i.e., $\lambda > 0.9$ and $T_\text{c} > 40$~K.

\begin{figure}[t]
	\includegraphics[width=1\columnwidth]{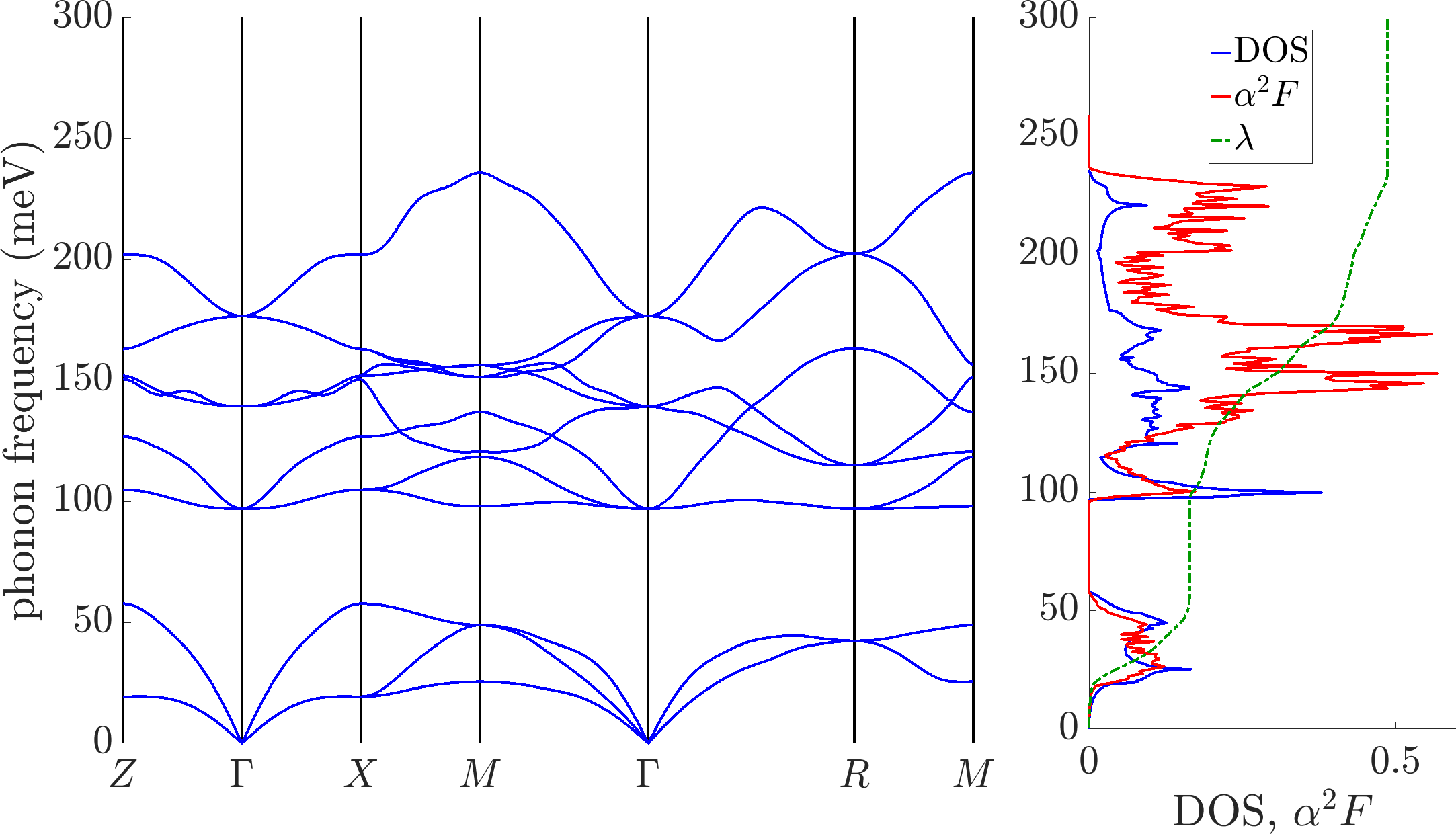}
	\caption{Phonon dispersion, phonon DOS, $\alpha^2 F$, and integrated $\lambda$ for NiH$_3$.} \label{fig:NiH3_phonons}
\end{figure}

\section{Crystal structures}
\label{app:crystal_struct}
In Table~\ref{tab:crystal_structures}, we report the crystal structures of all compounds mentioned in the main text.

\section{Vibrational and electron-phonon properties of NiH$_3$ and CoH$_5$}
\label{app:NiH3_CoH5}
\begin{figure}[t]
	\includegraphics[width=1\columnwidth]{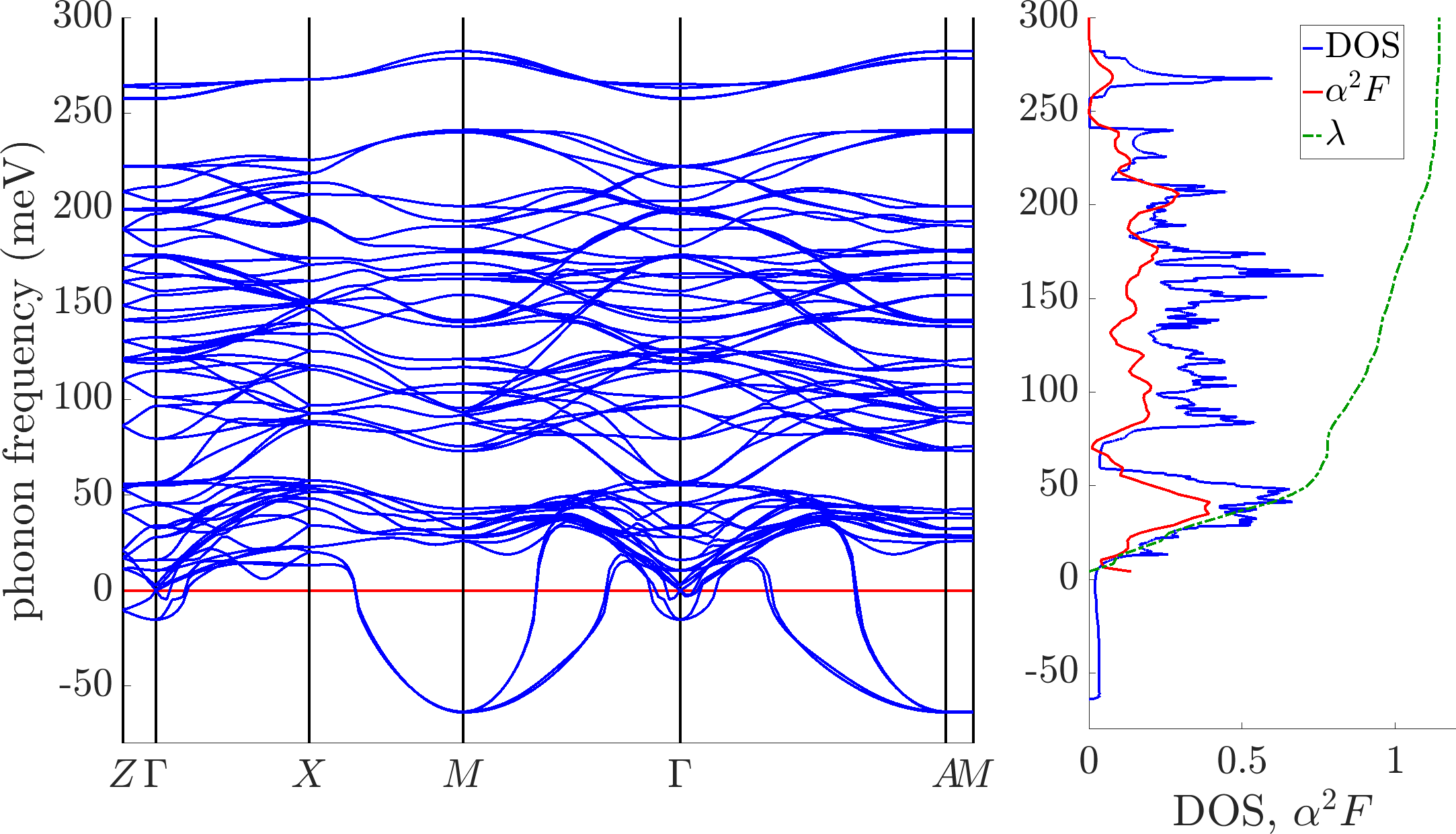}
	\caption{Phonon dispersion, phonon DOS, $\alpha^2 F$, and integrated $\lambda$ for CoH$_5$.} \label{fig:CoH5_phonons}
\end{figure}

Figures~\ref{fig:NiH3_phonons} and \ref{fig:CoH5_phonons} show the phonon dispersion, phonon DOS, $\alpha^2 F$, and integrated $\lambda$ for NiH$_3$ and CoH$_5$, respectively. Due to the computational expense, the vibrational properties for CoH$_5$ have been calculated only with \textsc{quantum espresso} on a $4 \times 4 \times 1$ Brillouin-zone grid in the 4-f.u. cell.

%\clearpage
\FloatBarrier

\bibliographystyle{apsrev4-1}

\end{document}